\pdfoutput=1
\documentclass[11pt]{article}
\usepackage{graphicx}
\usepackage{amsmath}
\usepackage{amssymb}
\usepackage{amsxtra}

\title{Towards a Derivation of Space}
\author{H. B. Nielsen${}^1$\thanks{e-mail: hbech@nbi.dk} and A. Kleppe${}^2$\thanks{e-mail:
    astri.snofrix@org}\\
${}^1$The Niels Bohr Institute, Copenhagen,
Denmark\\ 
${}^2$SACT, Oslo, Norway}
 
\begin{document}

\maketitle

\begin{abstract}
This attempt to ``derive'' space is part of the Random Dynamics project \cite{rd}. The Random Dynamics philosophy is that what we observe at our low energy level can be interpreted as some Taylor tail of the physics taking place at a higher energy level, and all the concepts like numbers, space, symmetry, as well as the known physical laws, emerge from a ``fundamental world machinery'' being a most general, random mathematical structure.
Here we concentrate on obtaining spacetime in such a Random Dynamics way.
Because of quantum mechanics, we get space identified with about half the dimension of the phase space of a very extended wave packet, which we call "the Snake".
In the last section we also explain locality from diffeomorphism symmetry.
\end{abstract}


\section{The space manifold}
This is an attempt to ``derive'' space from very general assumptions:

1) First we postulate the existence of a phase space or state space, which is quite general and abstract. It is so to speak an ``existence space'', with very general properties, and to postulate it is close to assume nothing. 

So we start with the quantized phase space of very general analytical mechanics:
{\vspace{0.3cm}}

\begin{displaymath}
\left\{
\begin{array}{ll}
q_1, q_2,...,q_N                                                                     \\ 
p_1, p_2,...,p_N = i\frac{\partial}{\partial q_1},...,i\frac{\partial}{\partial q_N}  \\ 
H(\vec{q},\vec{p})                                                                       
\end{array}\right.
\end{displaymath}

where $N$ is huge.
This is (almost) only quantum mechanics of a system with a classical analogue, which is a very mild assumption.

{\vspace{0.3cm}}
2) For the Hamiltonian $H$ we then examine the statistically expected ``random $H(\vec{q},\vec{p})$'' functional form (random and generic).

{\vspace{0.3cm}}
3) In the phase space we single out an ``important state'' and its neighbourhood - the ``important state'' supposedly being the ground state of the system.

{\vspace{0.3cm}}
The guess is that the ``important state'' is such that the state of the Universe is in the neighbourhood of this ``important state'' - which presumably is the vacuum.

\includegraphics[scale=0.5]{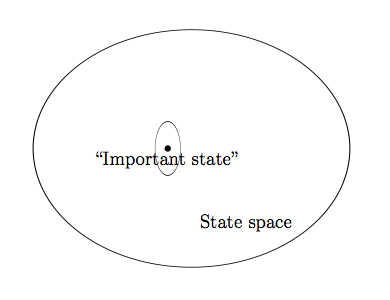}
The state we know from astronomical observations is very close to vacuum. According to quantum field theory this means a state which mainly consists of filled Dirac seas, with only very few true particles above the Dirac seas, and very few holes. This vacuum is our ``important state'', supposedly given by a wave packet.
If the system considered is the whole Universe, each point in the phase space is a state of the world.

Classically, a state is represented as a $point$ in phase space, but quantum mechanically, due to Heisenberg, this phase space point extends to a volume $h^N$. Now assume that this volume is not nicely rounded, but stretched out in some phase space directions, and compressed in others.

The phase space has $2N$ dimensions, so a wave packet apriori fills a $2N$-dimensional region. Our assuption is that the vacuum wave packet is narrow in roughly $N$ of these dimensions. The vacuum state is thus extended to a very long and narrow surface of dimension $N$ in the phase space (where $N$ is half the phase space dimension).

The really non-empty information in this assumption is that some of the widths are much smaller than others. $N$ is moreover enormous, equal to the number of degrees of freedom of the Universe, so our model is really like a particle in $N$ dimensions, $(q_1,q_2,...,q_N)$.  The ``important state'' is one where ``the particle'' is in a superposition of being in enormously many places (and velocities). 

We envisage the points along the narrow, infinitely thin wave packet as embedded in the phase space, and that they in reality are our space points. In relation to this infinitely narrow ``snake'', these points are seemingly ``big'' (one can imagine the points as almost 'filling up' the Snake volume in the transversal direction).
In the simplest scheme half of the phase space dimensions are narrow on this Snake, and the other half are very extended, long dimensions on the Snake.
  \begin{figure}[htb]
    \begin{center}
    \includegraphics[scale=0.5]{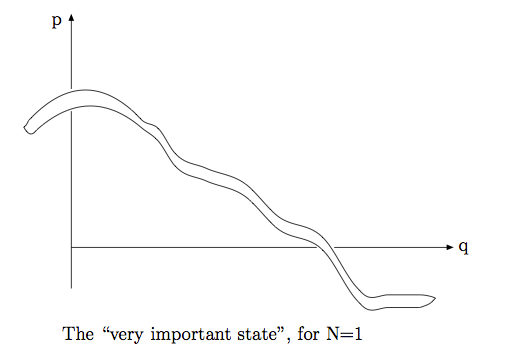}
\end{center}
    \end{figure} Along the Snake surface, the ``important state'' vacuum wave packet, i.e.
    the wave function $\Psi(q_1,q_2,...,q_N)$ of the Universe, is supposed to be
    approximately constant. With
    $\Psi \approx$ constant, reparametrization (once it has been defined) under
    continuous reshuffling of the ``points'' along the long directions of the
    wave packet, is a symmetry of the ``important state''.
\begin{figure}[htb]
    \begin{center}
    \includegraphics[scale=0.8]{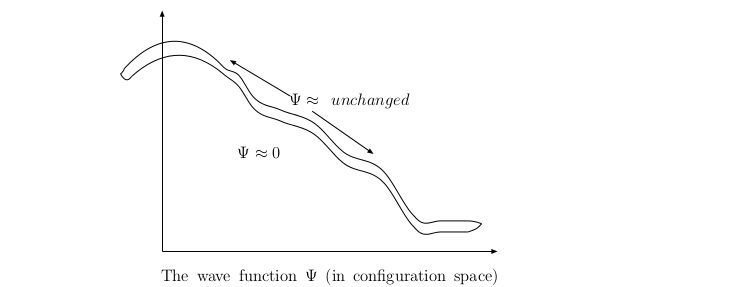}
\end{center}
    \end{figure}  
The idea is to first parametrize the $N$ ``longitudinal'' dimensions so $\Psi$ gets normalized to be $1$ all along the Snake.
It is however not $\Psi$ we are most interested in, but the probability of the Universe to be at x, corresponding to 
\begin{equation} 
\int_t |\Psi(x,y)|^2\mathrm{d}^N_ty
\end{equation} 
where $t$ stands for transverse. 

With some smoothness assumptions, the longitudinal dimensions will be like a manifold, i.e. the points given by the longitudinal dimensions constitute a ``space manifold''.
Since $N$ is huge, the wave packet extension is probably also huge. And since there is a huge number of possibilities in phase space, the Snake is most certainly also very curled. 

A wave packet can be perceived as easily excitable displacements of the transversal directions of the $N$-dimensional Snake (approximate) manifold. 
There are presumably different $q_i$ and $p_i$ at different points on the manifold, and states neighbouring to the vacuum (``the important state'') correspond to wave packets just a tiny bit displaced from the
vacuum. Thus the true state is only somewhat different from the vacuum (there is a topology on the phase space, so ``sameness'' and ``near sameness'' can be meaningfully defined).
Corresponding to different points on the long directions of the
wave packet (manifold), ``easy'' excitations can then be represented as some
combinations $\sum_i (\alpha_i \Delta q_i + \beta_i \Delta p_i)$ of the ordered set 
$(\Delta q_1,...,\Delta q_N,\Delta p_1,...,\Delta p_N)$, where $q_i$ and $p_i$ are different phase space points of the
$N$-dimensional manifold.
The ``easy'' degrees of freedom are thus assigned to points on the
manifold, so an ``easy'' displacement on the Snake is extended
over some region along the Snake, that is, in x. In that sense the
``easy'' degrees of freedom can be interpreted as functions of x,
$\phi_1(x)$, $\phi_2(x)$,...., which actually look like fields on
the manifold (this is just notation, but in some limit it is justified). The wave packet $\Psi$ consisting of easily
excitable displacements, can then be perceived as superpositions
of the $\phi_i(x)$.
A field is just degrees of freedom expressed as a function of x (a field actually has to be a degree of freedom, in the sense that it is among parameters describing the state of the Universe), and these superpositions really seem to be fields.  

Now, let us make superpositions of such ``easy'' displacements to form one only non-zero displacement very locally, this is certainly legitimate.
But with the identification of the Snake with space (or the space manifold), we should require that changing a field $\phi(x)$ only at $x_0$ corresponds to keeping the Snake unchanged, except at $x_0$.

So far we have identified the ``important state'' as the ``ground state'', i.e. the classical ground state $\approx$ Snake.
Now
consider the classical approximation for directions transverse to the Snake: In the transverse directions ($\sim$ $y$), taking $H$ as function of $y$ at the minimum of the crossing point with the Snake (chosen to be the origin), the Taylor expansion of $H$ with regard to $y$ near the Snake is given by (discarding unimportant constants) second order expansions
\begin{equation}
H \approx \frac{1}{2} \frac{\partial^2}{\partial y_i \partial y_j} H(y,x)|_{y=0} \cdot    y_i y_j
\end{equation}
We now diagonalize, i.e. look for eigenvalues of the matrix $(\partial^2 H/\partial y_i \partial y_j)_{ij}$, where the ``easy modes'' correspond to the lowest eigenvalues.

From smoothness considerations these eigenvalues $\omega_1, \omega_2,...$ can be defined as continuous and differentiable as functions of x, where x are the coordinates along the Snake. So, if $N \textgreater 3$, we could strictly speaking identify these eigenvalues by enumeration: The lowest, next lowest, etc., except for crossings.
As an example take a very specific Hamiltonian 
 \begin{figure}[htb]
    \begin{center}
    \includegraphics[scale=0.5]{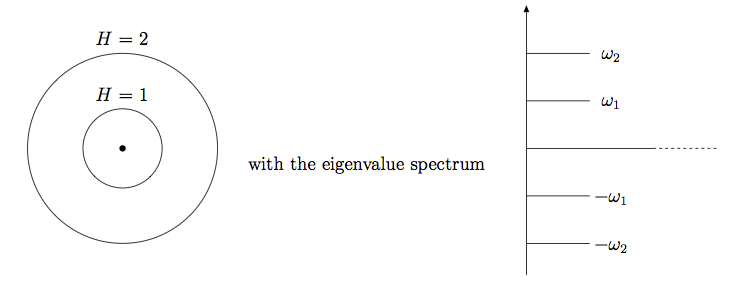}
\end{center}
    \end{figure} giving cotecurves of $H$ by choice of coordinates $y$, so $H \sim {\bar{y}}^2$, and the commutator [$y_i,y_j$] being very complicated.

\subsection{The vacuum Snake}

Until now, our main assumption is that the world is in a state in the neighbourhood of ``the vacuum Snake''.
The true Snake is in reality a state that can be considered
a superposition of a huge number of states that are all needed to
be there in the ground state because there are terms in the Hamiltonian
with matrix elements between these states (of which it is superposed).
We could think of these terms enforcing the superposition for
the ground state as some kind of ``generalized exchange forces.''
To go far away from the Snake would be so rare and so expensive that it in principle doesn't occur, except at the Big Bang.
It is also possible that the Snake is the result of some
Hubble expansion-like development just shortly after Big Bang.
It must in reality be the expansion that has somehow brought
the Universe to be near an effective ground state or vacuum, because
we know phenomenologically from usual cosmological models that
the very low energy density reached is due to the Hubble expansion.
Thinking of some region following the Hubble expansion, its space expands but we can nevertheless consider analytical mechanical systems. 
 \begin{figure}[htb]
    \begin{center}
    \includegraphics[scale=0.5]{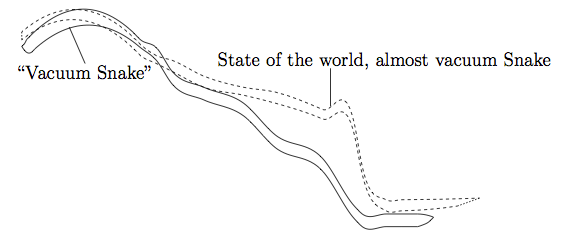}
\end{center}
    \end{figure} Starting with a high energy density state, i.e. rather far from vacuum, the part of the Snake neighbourhood which is used gets smaller and smaller after Big Bang. Already very close to the singularity - if there were one - the only states were near the Snake. 
We may get away from the ``Snake valley'', but only at Planck scale energies.
And we will probably never have accelerators bringing the state very far away from the Snake. 
So far, we have identified ``the Snake'' in the phase space of the very general and very complicated analytical mechanics system quantized.

Aiming at deriving a three-dimensional space, we must 
have in mind that this manifold, which is the protospace,
has a very high dimension of order $N$ which is the number of degrees
of freedom of the whole universe. If that were what really
showed up as the dimension of space predicted by our picture,
then of course our picture would be immediately killed by
comparison with experiment. If there shall be any hope for
ever getting our ideas to fit experiment, then we must at least
be able to speculate or dream that somehow the effective spatial
dimension could be reduced to become 3. 

For many different reasons, it seems justified to believe that 3 is the dimension of space. The naive argument is that we experience space as 3-dimensional, the number of dimensions is however not to be taken for granted, as we know from e. g. Kaluza-Klein, and string theory.
We shall in the following
at least refer to some older ideas that could make such a reduction
possible. For instance one can have that in some generic equations
of motion one gets for the particle only non-zero velocity in
three of the a priori possibly many dimensions.

\section{The number of space dimensions}

In the 1920-ies Paul Ehrenfest \cite{Ehrenfest} argued that for a $d$ = $D+1$-dimensional spacetime with $D > 3$, a planet's orbit around its sun cannot remain stable, and likewise for a star's orbit around the center of its galaxy. 
About the same time, in 1922,
Hermann Weyl \cite{Weyl} stated that Maxwell's theory of electromagnetism only works for $d = 3+1$ , and this fact \textit{"...not only leads to a deeper understanding of Maxwell's theory, but also of the fact that the world is four dimensional, which has hitherto always been accepted as merely 'accidental,' become intelligible through it." }
 
The intuition that four dimensions are 'special' is also supported by mathematician 
Simon Donaldson \cite{Donald}, whose work from the early 1980-ies on the classification topological four-manifolds indicates that the most complex geometry and topology is found in four dimensions, in that only in four dimensions do exotic manifolds exist, i.e. 4-dimensional differentiable manifolds which are topologically but not differentiably equivalent to the standard Euclidean $R^4$.

The existence of such wealth in 4-dimensional complexity is reminiscent of Leibniz' idea \cite{Leibniz} that God maximizes the variety, diversity and richness of the world, at the same time as he minimizes the complexity of the set of ideas that determine the world, namely the laws of nature. Only, Leibniz never told in what dimensions this should be the case, but according to Donaldson, this wealth of structure is maximal precisely in a 4-dimensional spacetime manifold.

\subsection{3+1 dimensions and the Weyl equation}

Another way to ``derive'' $3+1$ dimensions, is by assigning primacy to the Weyl equation \cite{hbech}. 
The argument is that in a non-Lorentz invariant world, the Weyl equation in $d=3+1$ dimensions requires less finetuning than other equations. This means that in $3+1$ dimensions the Weyl equation is especially stable, in the sense that even if general, non-Lorentz invariant terms are added, the Weyl equation is regained. So in this scheme both $3+1$ dimensions and Lorentz invariance eventually emerge.

Before $3+1$ dimensions there is no geometry. 
Starting with an abstract mathematical space with hermitian operators ${\bar{\sigma}}$ and ${\bar{p}}\psi$, and a wave function $\psi$ in a world without geometry, 
choose a two-component wave function,

\begin{equation}\label{nambu}
 {\bar{\sigma}}{\bar{p}}
              \left (\begin{array}{rcl}
                  \psi_1 \nonumber\\
                  \psi_2   
                      \end{array}
                \right) = 
         p_0\left (\begin{array}{rcl}
                  \psi_1 \nonumber\\
                  \psi_2   
                      \end{array}
                \right)
\end{equation}
where $p_0$ is the energy. In vielbein formulation this is $V_a^{\mu} \sigma ^a p_{\mu} \psi = 0$, which is the Weyl equation with hermitian matrices $\sigma^a$ that are the Pauli matrices $\sigma^1$, $\sigma^2$, $\sigma^3$. The vielbeins are really just coefficients coming about because we write the most general equation.
The Weyl equation is Lorentz invariant and the most general stable equation with a given number of $\psi$-components, and as a general linear equation with $2{\rm { x}} 2$ hermitian matrices, it points to $3+1$ .

In $d$ dimensions the Weyl equation reads
\begin{equation}
\sigma^a e^{\mu}_a \frac{\partial \psi}{\partial x^{\mu}}=0,
\end{equation}
$a$=(0,1,2,3), and
the metric
$g^{\mu\nu}=\sum\limits_{a} \eta_{aa} e^{\mu}_a e^{\nu}_a
$
is of rank=4. If the dimension $d > 4$, there is however degeneracy.

For each fermion, there are generically two Weyl components. 
If we had a generic equation with a 3-component $\psi$, we would in the neighbourhood of a degeneracy point in momentum space, have infinitely many points with two of the three being degenerate.

Assume that $\psi$ has $N$ components, $\psi = (\psi_1,....,\psi_N)$. Consider a $C$-dimensional subspace of the $\psi$-space spanned by the $\psi$-components $\psi_1,...,\psi_C$, 
with $N \geq C$, and at the ``$C$-degenerate point'', there is a $C$-dimensional subspace in $\psi$-space ($N$-dim) for which $H\psi=\omega\psi$, with only one $\omega$ for the whole $C$-dimensional subspace (degenerate eigenvalue $\omega$ with degeneracy $C$ - the eigenvalue $\omega$ is constant in the entire $C$-dimensional subspace).
 \begin{figure}[htb]
    \begin{center}
    \includegraphics[scale=0.4]{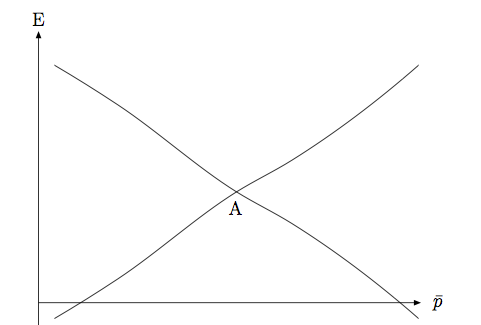}
\end{center}
    \end{figure}
In the neighbourhood we generically have $\bar{p} \bar{\gamma}$ extra in $H$, where 
\begin{equation}
H(\bar{p})=H(\bar{p}_{degenerate})+\bar{p}_0\bar{\gamma}
\end{equation}

for which $H\psi=\omega \psi$.
There are lower degeneracy points in the neighbourhood (meaning $p^{\mu}$-combinations with more than one polarization), 
where in the situation with two polarizations. In the above figure $A$ represents the 2-generate point and the curves outside of $A$ represent the situation where only one eigenvector in $\psi$-space is not degenerate.
In the neighbourhood of a ``generic'' 3-degenerate (or more) point there are also 2-degenerate points.
But the crux is the filling of the Dirac-sea.
Think of the dispersion relation as a topological space:
Can we divide this topological space into two pieces, one ``filled'' and one ``unfilled'' so that the border surface $\partial``unfilled''=\partial``filled''$ only consists of degenerate states/dispersion points?
If not, we have a ``metal''. 

The question is whether there is a no-metal theorem. To begin with, we can formulate one almost trivial theorem:
If the border $\partial``unfilled''$ contains a more than 3-degenerate point, we generally either have a metal or else 2-degenerate points on this border.
There is also the disconnected dispersion relation, corresponding to an insulator.

Counter example: Imagine a 6-dimensional Weyl equation. In this case, the border $\partial``filled''$ has only one point in the 6-dimensional Weyl, so there is only a 4-degenerate point and no 2-degenerate points on the border. 
 \begin{figure}[htb]
    \begin{center}
    \includegraphics[scale=0.8]{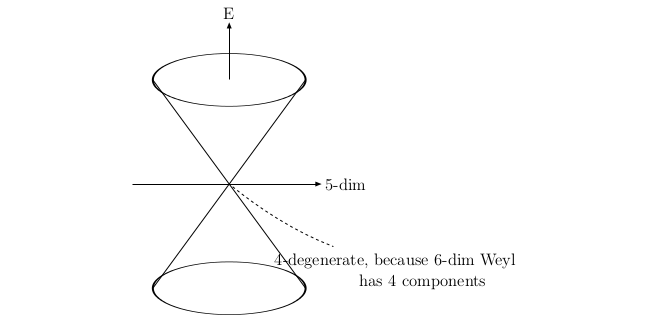}
\end{center}
    \end{figure} The statement about the stability of the Weyl equation in $3+1$ dimensions would thus be false if the 6-dim Weyl were ``generic''. But it is not, so there is no problem.

In $d$ dimensions the number of $(1+\gamma^5)\gamma^{\mu}$ matrices is $d$ (where $(1+\gamma^5)$ project to Weyl, i.e. the handedness), and the Weyl $\psi$ has $2^{d/2-1}$ components. That means that there are $2^{d-2}$ matrix elements in each $(1+\gamma^5)$ projected $\gamma^{\mu}$. Assuming that the dimension $d$ is even, normal matrices $\gamma^{\mu}$ (i.e. Dirac gamma matrices) have $2^{\frac{d}{2}}$ matrix elements in each $\gamma^{\mu}$.

Now, for $2^{d-2}\hspace{2mm} \textgreater\hspace{2mm} d$, one can form matrices which on the one hand act on the Weyl field $\psi$ (with its $2^{\frac{d}{2}-1}$ components), but on the other hand are not in the space spanned by the projected $\gamma^{\mu}$-matrices. One could in other words change the Weyl equation by adding some of these matrices, thus for $2^{d-2}\hspace{2mm} \textgreater\hspace{2mm} d$ the Weyl equation is not stable under addition of further terms. 
So the Weyl equation is not ``generic'' for $2^{d-2}\hspace{2mm} \textgreater\hspace{2mm} d$, i. e. it so to speak has zero measure (in the sense that if you write down a random equation of the form $[\sum\limits_{a} p_a M^a(\mathrm{n x n})]\psi =0$ in $d$ dimensions, where $n$ is the number of $\psi$-components and $2^{d-2}\hspace{2mm} \textgreater\hspace{2mm} d$, the probability that it is the Weyl equation is zero).
It is on the other hand impossible to have $d$ linearly independent projected $\gamma^{\mu}$-matrices if $2^{d-2}\hspace{2mm} \textless\hspace{2mm} d$, for even dimension $d$. 

Looking at different number of dimensions $d$, we conclude that for $d = 4$, $2^{d-2} = d$, seemingly confirming the ``experimental'' number of dimensions $4= 3+1$, i.e. there is genericness: It seems like the 4-dimensional Weyl equation is just the most general stable equation with a given number of $\psi$-components.  

\begin{center}
    \begin{tabular}{| l | l |}
    \hline
     $d$ & \hspace{5mm} $2^{d-2}$    \\ \hline
     0 & \hspace{6mm} 1/4 \\ \hline
     1 & \hspace{6mm} 1/2 \\ \hline
     2 & \hspace{8mm} 1 \\ \hline
     3 & \hspace{8mm} 2             \\ \hline
     4 & 4 - equality! \\ \hline
     5 & \hspace{8mm} 8             \\ \hline
     6 &\hspace{7mm} 16             \\ 
    \hline
    \end{tabular}
\end{center}
So on the one hand the experienced number of dimension is $4 = 3+1$ , and on the other hand, in $d$ = 4 the Weyl equation is stable under small modifications (so here the Weyl equation is ``generic'').

\subsection{Bosons and fermions}
Arguing that space has $3+1$ dimensions, we however run into
the old story that we get $3+1$ dimensions and Lorentz invariance separately for each type of particle.

From one perspective, fermions should however not exist at a fundamental level, since they violate locality, 
\begin{equation}  
[{\bf{\psi}}(\bar{x}),{\bf{\psi}}(\bar{y})]\neq 0
\end{equation}
One way out could be to get effective fermions from bosons, \`{a} la the relation in 1+1 dimensions,
\begin{equation}
{\bf{\psi}} \sim_{def} e^{i \phi} 
\end{equation}
where $\phi$ is a boson field.
If there are $N_f$ fermion components and $N_b$ boson components, then moreover \cite{HolgerAratyn}
\begin{equation}
\frac{N_f}{N_b}\approx \frac{2^{d-1}}{2^{d-1}-1}
\end{equation}

A bosonic counterpart to the Weyl equation would be of the form
\begin{equation}\label{boson}
K^{\mu}_{ba}\partial_{\mu}\psi_a=0,\\ 
g^{\mu\nu} = K^{\mu}_{ba}K^{\nu}_{cd}\Pi^{bacd}
\end{equation}
 where e.g. $\Pi^{bacd}$=$\delta^{ba}\delta^{cd}$, and $K^0 = \delta^{ab}$ for $a=b$, and $K^i_{ba}=i\epsilon^i_{ab}$, $H=1/2\sum\tilde{\psi}_a^2(\bar{p}) \rightarrow \delta_{ab}\tilde{\psi}_a\tilde{\psi}_b$.

In the game for gauge bosons or Weyl fermions, we look for a mechanism of aligning the metrics for the different species of particles. We want to generalize the coherent state concept and show that the states on the manifold can be called generalized coherent state. 
Coherent states are usually given from harmonic oscillators with $q's$ and $p's$. 
So we must locally (in the phase space) approximate the system by harmonic oscillators, then seek to extract $q's$ and $p's$ as operators, and so we might have proven the quantized analytical mechanics model. 

Define a generalized coherent state 
$A(q,p)q_{op} + i B(q,p)p_{op}$,
such states are given by points on a manifold. Differentiating with respect to a coordinate on the manifold should give $p$ or $q$ acting on the state,  
\begin{equation}
(Aq_{op} + i Bp_{op})|q',p'>=(Aq' + i Bp')|q',p'>
\end{equation}

One thing is to have a manifold of rays, another is to have one of state vectors (in the Hilbert space)
$|\lambda> = e^{\lambda a^{\dagger}}|0>$,
\begin{equation}
\frac{d}{d\lambda}|\lambda>\approx a^{\dagger}|\lambda>
\end{equation}
$a^{\dagger}= \alpha q+ip$

As point of departure, we use gauge particles at low energy.
There come metrics out of it, one for each gauge boson.
The equation of motion we get is 
\begin{equation}\label{frog}
\partial_t \left(
\begin{array}{c}
\phi_1\\
\phi_2\\
\phi_3\\
\end{array}
\right)=
i \left(\begin{array}{ccc}
0       & A_{12}  & A_{13} \\
-A_{12}  & 0      & A_{23} \\
-A_{13}  & -A_{23} & 0      \end{array} \right)
\left(
\begin{array}{c}
\phi_1\\
\phi_2\\
\phi_3\\
\end{array}
\right)
\end{equation}
where 

${\bf{A}} \approx \bar{p}\hspace{2mm}\rm{and}\hspace{2mm} \phi_i=B_i+i E_i \simeq F_{jk}\epsilon^{jk}_i+iF_{0i}$.

Together with C. Froggatt, one of us has 
shown \cite{FroggatBechNielsen} that looking at the very low energy behavior of a (rather) generic
system of bosons, one may arrive at an approximate equation of
motion for three of the fields of the form (\ref{frog}). However, typically for Random Dynamics,
we should argue that the coefficients the A's here are dynamical.
These A's are (essentially) the same as the K's in equation (\ref{boson})
and we have already written that a metric tensor comes out of them. Of course all fields are basically of the form of some combination of the $\phi_i(x)$'s, since they make up
at least all the ``important'' degrees of freedom. This is also true for the A's, or equivalently
the K's, thus in the end the metric tensor comes to depend on the $\phi$'s.

\section{Reparametrization}
If a space has $N$ dimensions, the phase space dimension is 2$N$, and the Hilbert space can be perceived as a sum ${\cal{H}}=\sum_{\oplus}{\cal{H}}_N$. $N$ is not a constant of the motion, so we need some term in the Hamiltonian going from one $N$ to another. So let us imagine an only quantum mechanically describable term
with matrix elements between wave packets connected to the phase space
for one $N$, and the wave functions connected to another of the $N$ values
(another phase space so to speak).

The full Snake must then be imagined as really a superposition of
one (or more) snakes in each or at least several of the phase spaces corresponding to the various $N$ values. Hereby the snakes in the different
$N$-value phase spaces get locked together, but they will somehow be locked
so as to follow each other - due to the quantum matrix elements connecting the different $N$-value phase spaces - and we effectively have only one snake. 

We let x enumerate the points along the Snake, i.e. in the ``longitudinal direction'', x is chosen by convention. We can just as well choose again, now choosing it to be x'=x'(x), it should not matter. The crux is whether the action is independent on these choices, i.e. whether $S(\psi_i(x),...)$ and $S'(\psi_i(x'),...)$ are of the same form, supposedly something like 
\begin{equation}
S=\int(\displaystyle\sum_{i} \dot{q}_ip_i-H)dt,
\end{equation}
presumably they are not. That means that reparametrization invariance is not automatically given, but must be derived.

In General Relativity we have $S=\int R \sqrt{g} \mathrm{d}^4x$. If we put $x'^{\mu}=x'^{\mu}(x^{\rho})$ into $S$, and transform $g_{\mu \nu}$the conventional way, $g'_{\mu \nu}(x')$=$g'_{\mu \nu}(x') (..)$, we get $S=S'$ from the constructed form (of Einstein and Hilbert).
But since we have no a priori reparametrization invariance, we cannot state that the action is independent in this way. So far, our Snake model doesn't even have translational invariance. It needs to be derived, and we also need to derive diffeomorphism invariance. 

Following the scheme of Lehto-Nielsen-Ninomiya \cite{NN}, the diffeomorphism invariance should be achieved by quantum fluctuations, in the sense that quantum fluctuations should produce translational invariance and in the end even reparametrization invariance.

We do this by relating points on the Snake to points 'on' the metric (assuming that the effects of going along the string on the effective parameters that are being averaged are bounded, so that the average at least converge):
Consider a point given by computation using the $g^{\mu \nu}$, which quantum fluctuates. These fluctuations so to speak smear out the differences between points chosen on the Snake, thus ensuring translational invariance (and diffeomorphism invariance).

In this way we can always formally get diffeomorphism invariance, but we risk to have some absolute coordinates functioning as ``Guendelman variables'' \cite{Guen}. To show practical reparametrization invariance then depends on how we get rid of these absolute coordinates, or rather how their effects are washed away.

\subsection{Procedure}

1. We have the Snake in the phase space of the very general and very complicated analytical mechanics system quantized. 
  \begin{figure}[htb]
    \begin{center}
    \includegraphics[scale=0.8]{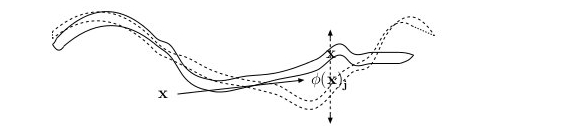}
\end{center}
    \end{figure}
We get the fields $\phi_j$ corresponding to the small displacements in transverse directions in which the frequencies of vibrations are ``small'' at the position x, telling where in phase space we are in the longitudinal directions of the Snake.

2. We assume (or show) that there are some fields (essentially among the $\phi_j(x)$'s or related to their development), a set of ``upper index metric fields'' $g^{\mu\nu}(x)$.

As a matrix, this metric should have rank 4, and we expect to find one $g^{\mu\nu}$ for each species of particles. Here we first think of gauge particles, postponing the fermions.

That is to say, we get some equations of motion for three effectively relevant fields $\phi_m$ (m=1,2,3) for each gauge particle species.

With equation (\ref{boson}) in mind, we consider the form
\begin{equation}
K^0_{mn}\tilde{\phi}_n-K^j_{mn}p_j\tilde{\phi}_n=0
\end{equation}
or just $K^{\mu}_{mn}p_{\mu}\tilde{\phi}_n=0$, where 
the $p^{\mu}$ stands for 
$p^{\mu} - p^{\mu}_0$, and $p^{\mu} = i \partial / \partial x^{\mu}$. 
\begin{equation}
K^0_{mn}=\delta_{mn} \hspace{5mm}{\rm{and}}\hspace{5mm}
K^i_{mn}=i\epsilon^{imn}
\end{equation}
But at first we only have 
\begin{equation}
K^{\mu}_{mn}=K^{\mu *}_{nm}  \hspace{5mm}({\rm{hermiticity}})
\hspace{5mm}{\rm{and}}
\hspace{5mm}K^0_{mn}=\delta_{mn}\hspace{5mm}({\rm{essentially\hspace{2mm} definition}})
\end{equation}
because we have chosen the simple Hamiltonian
\begin{equation}
H= \int (\sum_m{\tilde \phi_m}^2(\vec{p} ))d^{d-1} \vec{p}
\end{equation}
to be $\delta^{mn}{\tilde\phi_n}{\tilde\phi_m}$, and $K^i_{mn}=-K^i_{nm}$, because we let all the $K^i_{mn}$ come from the Poisson bracket (commutator) 
\begin{equation}
[{\tilde\phi_m}(\vec{p}),{\tilde\phi_n}(\vec{p})]=K^i_{mn}(p'_i-p_{i0})
\end{equation}
near the zero point in $\vec{p}$-space. From this $K^{\mu}_{mn}$ one then constructs the defining relation
\begin{equation}
g^{\mu\nu}=K^{\mu}_{mn}K^{\nu}_{op}\delta^{mo}\delta^{np}
\end{equation}
for the rank 4 metric with upper indices $g^{\mu\nu}$.

3. Assume (this must be true) that what we conceive as a point in space is calculated by using a metric $g^{\mu\nu}$ (we may have the problem of getting too many matrices $g^{\mu \nu}$, i. e. $g^{\mu \nu}_1(x), g^{\mu \nu}_2(x), g^{\mu \nu}_3(x),...$) integrating it roughly up to calculate where we have a point with given coordinates.

4. The formulation we shall use is by construction diffeomorphism invariant for the coordinate set x enumerating the points along the Snake. But that does $not$ mean that we have a diffeomorphic symmetric Hamiltonian $H$ or action $S$. We can namely have an underlying absolute coordinate system - or ``Guendelman variables''.
We could indeed imagine that we at first describe the longitudinal manifold along the Snake by a set of coordinates $\xi$, as many $\xi$ as there are x-coordinates, of course. When introducing the diffeomorphism transformable x, we perceive $\xi(x)$ as some (scalar) fields which are functions of x. But all the special structure of the phase space or analytical mechanics system as it varies along the Snake, appears as explicitly dependent on $H$, or $S$ on the $\xi$'s taking specific values. There is so to speak no translational invariance in $\xi$, but there is trivially in x, since translation is (apart from boundary problem) just a special diffeomorphism. Since in the ``vacuum'' it could at first seem that the $\xi$'s have in x varying values as one goes along in x, the presence of these $\xi$ (expectation) values even in ``vacuum'' means a spontaneous breakdown of translational invariance, and even more a spontaneous breakdown of diffeomorphism symmetry. 

At first glance, it thus looks like the ``Guendelman'' $\xi$-fields imply a spontaneous breakdown of translational and diffeomorphism invariance. So to prove that we do indeed have diffeomorphism invariance for say the Hamiltonian $H$, we must show that the practical effects of the ``Guendelman fields'' or original absolute coordinates $\xi$, wash out. Under the conditions which we shall consider, the $\xi$-dependent effects in practice average out. We shall argue that if we (as humans or physicists) count our position by integrating up some of the $g_{\mu \nu}$ obtained from $g^{\mu \nu}$ (or some average of $g^{\mu \nu}_1, g^{\mu \nu}_2, g^{\mu \nu}_3,..$), we fluctuate around relative to the $\xi$-coordinates (which are fixed in phase space along the Snake). These fluctuations were assumed under 3.

5. Now we need the assumption that the potentials, or more generally the Hamiltonian contributions depending on $\xi$ (and thus via the spontaneous breakdown violating the translational invariance), are bounded or at least as effectively bounded as fluctuations of $\xi$, so the averages over large regions in $\xi$ become (approximate) constants.

By taking this boundedness of the $\xi$-dependent part of the Hamiltonian as a reasonable assumption, the $\xi$-dependent contributions to the Hamiltonian wash completely out to nothing, the reason being the integrated up metric becomes integrated up over regions in x-space of the order of the size of the Universe, whereby the fluctuations become enormous.

If that is so, we have shown that for ``us'' situated in a place determined from the metric tensor fields $g_i^{\mu\nu}$ or rather their inverse $g_{\mu\nu}$ by long distance integration, the diffeomorphism invariance has been (effectively) (re)stored. In this way the formally introduced diffeomorphism invariance - just by thinking of x as an arbitrary set of variables - has become a good symmetry because of the $\xi$'s representing the lack of diffeomorphism symmetry by spontaneously breaking it, have gone practically out of the game.

It should be noticed that by this argumentation we have argued for diffeomorphism symmetry in the whole x-space of dimensions suspected to be 3, even if the metric tensor only has (because, say, of inheriting from $K^{\mu}_{mn}$) rank 4, thus delivering an effective spacetime of dimension $3+1$ . 

The point is that even though the single $g^{\mu\nu}(x)$ has only rank $4 = 3+1$ , it can fluctuate so all the fluctuation values of $g^{\mu\nu}(x)$ are included, and all directions in x-space covered. One may imagine the 3-dimensional space as a 3-dimensional submanifold embedded in the much higher dimensional x-space (the longitudinal space on the Snake). Then this submanifold not only fluctuates by extending and contracting in its own 3-dimensional directions, but also fluctuates around its transverse directions inside the x-space. Thus by quantum fluctuations (integrated up), the 3-space submanifold floats around (almost) all over the Snake in its longitudinal space.

For each fixed configuration of $g^{\mu\nu}(x)$ one has a whole ``fibration'' of 3-spaces lying parallel to each other in the x-space. 
Then the whole fibration fluctuates around in x-space. Accepting the above, we arrive at an approximate Hamiltonian (or an approximate action $S$) being exactly diffeomorphism invariant, whereby we can deduce locality.
After having derived locality that way, we get a picture very close to a model with gauge bosons and a dynamical metric, seemingly with 3 space dimensions. It looks rather like what we see phenomenologically, but there are a few weak points:

\begin{itemize}
\item The problem of each particle species, here each gauge particle species, having its own $K^{\mu}_{mn}(x)$, and thus it own $g^{\mu\nu}(x)$.

\item We have in some sense much more than 3 spatial dimensions because we have as many as has the longitudinal direction on the Snake.
\end{itemize}
These problems may not be very severe:
Calculating our position from the $g^{\mu\nu}$ as if space were 3-dimensional, we obtain what we use as position. Then it does not matter so much that relative to the Snake, the $\xi$-absolute coordinates fluctuate both in the 3 and the many other coordinates. Since the $\xi$'s are supposedly bounded - and thus relatively easy to average out to a constant -
it will just become even easier to get them averaged out over the bigger region where the position of ``us'' fluctuates.

We should however have in mind that signals going along the 3-dimensional surfaces along which the quanta can move, for every fixed imagined position of the 3-manifold inside
the much higher dimensional x-space, will only be able to move along that 3-surface. However, when this surface fluctuates wildly, also the signals running on it get swept along in much more than three directions.
That will however not be noticed by the physicist using the point-concept
resulting from integrating up the metric $g_{\mu\nu}$, or what we consider the more genuinely existing ({\tiny{$\therefore$}} a bit more fundamental)
$g^{\mu\nu }$. The physicist can only get motions in the three dimensions, simply because he only evaluates three coordinates in his position calculations.

So this problem is not so severe.

We however need to resolve the problem that each particle species has its own metric. A plausible solution goes in the direction that the metrics are in some way ``dynamical'',
and interact with each other in such a way
that they finally align, thus behaving as if they were all proportional
to each other. We would hope that e.g. the metric determining the gluon propagation would by interaction with the metric tensor (similarly related to say the W's and determining their propagation) bring them in the lowest energy situation
to become aligned, where this aligning then really should stand for that they become proportional to each other.

It should be noted that our theory is a priori not Lorentz invariant, at least not in the metric degrees of freedom, the Lorentz invariance supposed to be derived subsequently. 
Considering that our Snake is in its ground state, there are no ghosts, the question is how the different metrics behave. To begin with, we ask how $one$ metric $g_{\mu\nu}$ can avoid having ghosts. 

\subsection{Idea of Attracting Metric Tensors}

The basic idea in getting dynamical metrics which are adjusted to be parallel/proportional is not so difficult. Multi-metric gravity is however complicated by the (Boulware-Deser) ghosts \cite{BoulDes} that threaten to appear as one of the gravitons becomes massive.
Indeed lets us give the main hope:

1. For each type of particle, initially meaning each type of gauge particle (but if we add fermions
we could also have a metric tensor for each type of Weyl particle) there is a characteristic metric tensor $g^{\mu\nu}$ (with upper indices, prepared for being contracted with a derivative $\partial_{\mu}$ w.r.t. to the coordinates $x^{\mu}$). So we shall strictly speaking attach a particle species name to each of these metrics, e.g. $g^{\mu\nu}_{(W)}$ for the metric assigned to the $W$ gauge boson.

2. we argue that this metric is ``dynamical'' and even a field. Thus, it is not just a constant metric, but such that it

\begin{itemize}

\item can vary with initial conditions and fluctuate quantum mechanically,

\item can vary in time,

\item and even in space, since we take it as a field (we anyway have no translational invariance yet). The coefficients in the time development of the fields which are going to be interpreted as the gauge boson fields, will nevertheless depend on the precise position of the Snake
near the place to which the fields in question are assigned. The point
of view that the coefficients which give rise to the metric tensor are fields should be unavoidable.
\end{itemize}
3. Taking seriously the Random Dynamics assumption that everything that is allowed to interact also \textit{does} interact, we deduce that the different metric tensors associated with different particle species
will indeed interact.

4. We introduce the symmetries restricting the interactions between the various fields, paying attention to the metric
tensor fields associated with the different
particle species. At some point we get reparametrization from diffeomorphism invariance, which then restricts the way these metrics (which transform as upper index tensors) interact. Do not forget that by taking the inverse of the upper index matrix we can get one with lower indices instead (were it not for the problem that the metric only has rank 3
+1 and thus canot be inverted).

5. These restrictions from diffeomorphism or other symmetries,
also mean that the equations between the fields (resulting from the minimum energy state for the system w.r.t. to, say, the metric tensors)
also share these symmetries. This gives hope that the metric tensors will come to be proportional (or even equal) to each other.

6. Now, if the metric tensors for the different species of particles indeed get proportional, it really means that the Lagrangian terms or equations of motions for the different particle species can be written with the {\em same} metric and just some overall factors in addition. This in its turn means that in the end, there are no effectively different metrics.

If you have several different metrics, this is what supposedly happens:

\begin{itemize}

\item You get bigravity or multigravity,
meaning that you get a model with several
spin=2 particles \cite{Damour} \cite{Speziale}.

\item We can (after some partial gauge fixing) interpret the massless graviton
as a Nambu-Goldstone particle for diffeomorphism symmetry, and we expect that even after getting several metric tensors we should only have one massless graviton if the diffeomorphism symmetry remains \cite{Ogivetsky}.
So we expect one massless graviton and several massive spin 2 particles, namely
the number of metric tensors minus one.

The graviton becomes a real Nambu-Goldstone particle due to a linearly varying gauge function.
Simple shift by adding a constant to a coordinate, perceived as a reparametrization/gauge transformation, 
is not spontaneously broken in Einstein gravity. It's only the linear variation of $\epsilon$ with x, that makes the metric tensor field spontaneously breaking the transformation.

\item Then our ``poor physicist thinking''
means that we guess that all particle species which don't have a reason for being massless (or almost massless), have so big masses that they are in practice not present (it is so to speak the Universe after the very first singularity (supposing there was one), which is so cold that massive particles do not occur even if they exist in the sense that they could in principle be produced in some
enormously expensive accelerator).

This means that all the heavy graviton field degrees of freedom are in their no-excitation state. If these fields are the metrics, or better some linear combination of metrics for the different particles, the non-excitation of the majority of these linear combinations leaves only one excitable combination $\sum a_i g^{\mu\nu}_i =
a_W g^{\mu\nu}_W + a_{gluon}g^{\mu\nu}_{gluon}+ ...$
of the various metrics, namely the massless combination. This means that the various metric fields are forced to follow each other. They will namely all follow the massless graviton field, simply being equal to this massless metric multiplied by some constant.
\end{itemize}
If indeed a massive spin two graviton would appear, there will
no longer be any proportional metrics.
But that would be rare, and we would interpret the effect of having 
different metrics for different species as effects of interaction with this heavy graviton.

So once we have argued that the metric tenors are dynamical and interacting, there is really good hope for getting rid of the old problem in Random Dynamics, that different species have different metric tensors. The crux of the matter is that the different metrics have the chance to dynamically influence each other, and thereby for symmetry reason become (apart from some extra factors) the same metric.

\subsection{General Ghost Problems}

Making theories with one or several massive gravitons, i.e. bigravity, is highly non-trivial due to the ghost-problem of Boulware and Deser.
The problem is that if you essentially randomly create
theories for spin 2 particles, you are very likely to run into the problem of unstable modes of vibration.
We here think of classical fields, and

for the theory to be stable - i.e.
have a bottom in the Hamiltonian -
all modes of vibration should be like harmonic oscillators rather than
like $inverted$ harmonic oscillators.
It is, however, rather an art to avoid getting such ghosts or unstable vibrations, if one seeks a massive spin two. Thereby it becomes a problem also for making an interacting bigravity or multigravity. We argued that we expected only one massless graviton. If we have several, it is most likely that one or more are heavy gravitons, which then in turn brings their ghost-problem.

Hassan and Rosen \cite{Bigravity} argue that they have got the only bigravity without ghosts. A characteristic of this two metric theory (= bigravity) is that the interaction, apart from the usual factor $\sqrt{-det g}$, is a function only
of a kind of ratio of the interacting metrics $f_{\mu\nu}$ and $g_{\mu\nu}$,
formally written $\sqrt{g^{-1}f}$.
This means that it depends on a constructed metric $\gamma^{\mu}_{\nu}$
defined by the equation
\begin{equation}
\gamma^{\mu}_{\nu}\gamma^{\nu}_{\rho} = g^{\mu \rho}f_{\rho \nu}.
\end{equation} In fact the interaction part of the Lagrangian density is written as a sum with coefficients $\beta_n$ of symmetrized
products of eigenvalues of the matrix $\gamma^{\mu}_{\nu}$.

There is as a side remark for us who
have a theory in which the metric tensor appears as a product of two matrices: We may construct the square root
matrix  $\gamma^{\mu}_{\nu}$ directly from the matrices that must essentially be squared to obtain the metric, i.e. our original variables from which we construct the metric are already a kind of square roots of the metric.

Concerning the Bouleware-Deser ghosts or unstable modes, for the purpose of our machinery for obtaining
relativity and space, we may think as follows:

If we have chosen to consider states around a ground state which has the lowest possible energy, there cannot be any vibration modes unless the vibration leads to positive or at least non-negative energy. That means that all the vibrations around our ground state - the ground state of the Snake
- must be of the type of a positive frequency and energy, i.e. ordinary rather than inverted harmonic oscillator. So from our a priori very general model one deduces a good behavior of the resulting particle field equations.
There shall be no unstable modes of vibration in the effective field theory resulting from our Snake model. We logically allow a type of bigravity or massive gravity which avoids the ghosts, and if it is claimed that there is no alternative to a certain special type of models to avoid the instabilities (that would mean that the bottom falls out of the Hamiltonian, so some states would have energy less than the state assumed to have the lowest energy around which we expand) we should be formally allowed to conclude that this type of model is effectively realized in our Snake model.
It's only once we manage to get dynamical metrics that the discussion of bigravity type theories becomes relevant, but we at least get some coefficient-fields which we strongly expect to become dynamical variables. Surely there will to these fields, which if dynamical, formally correspond to some ``metric tensors''.

In the spirit that all allowed terms should be there, the speculation that these metric fields must obtain some kind of kinetic term in our very general model, seems very well supported.
This is essentially just the Random Dynamics assumption that the coupling parameters can be considered random, so they cannot be in any (simple) special value system that would have measure zero. Thus the possible kinetic terms must be allowed, and the sign(s) can only be as needed for the already discussed ground state to indeed be the ground state.

We take this
argumentation to mean that we must expect our very general analytical mechanical system treated as the Snake to be approximated by the matter gauge fields (and Weyl fermions if we allowed), in addition to a say in the two
gauge boson case (for simplicity) the bi-gravity of Hassan and Rosen, cleverly
adjusted to have no instabilities ({\tiny{$\therefore$}} no ghosts). This Hassan Rosen model should apart from possible modifications of the kinetic energy have an action like
\begin{equation}
S = M_p^2\int d^4x\left[ \sqrt{-g} \left (
R + 2m^2 \sum_{n=0}^4\beta_n e_n(
\sqrt{g^{-1}f})   \right) +M_{pf}^2 \sqrt{-f}
\right] \end{equation} which is equation (2.1) in \cite{Bigravity}
with a kinetic term $\propto R_f$ for the $f_{\mu\nu}$.
This equation looks a bit less symmetric than it will be in the end. The notation is that we have two metric tensors $g_{\mu\nu}$ and $f_{\mu\nu}$
and $R$ denotes the usual Einstein Hilbert action scalar curvature calculated
from $g_{\mu\nu}$ in the usual way. The symbols $g$ and $f$ are of course the determinants of the two metric fields, but the symbol $\sqrt{g^{-1}f}$
is \emph{not} related to the determinants but rather it means a \emph{matrix} $\gamma^{\mu}_{\nu}$ determined as the square root from the condition:
\begin{equation}
\gamma^{\mu}_{\nu}\gamma^{\nu}_{\rho}=
g^{\mu\nu}f_{\nu \rho}
\end{equation}
Notice the natural use of $g^{-1}$ for
$g^{\mu\nu}$ which is of course the inverse of the $g$ matrix $g_{\mu\nu}$
as the metric with upper indices always
is.

The symbols $e_n(\gamma^{\mu\nu})$
for $n$ running from 0 to 4 are the symmetrized eigenvalues of the matrix
$\gamma_{\mu\nu}$. That is to say
\begin{eqnarray}
e_0(\sqrt{g^{-1}f}) & = & 1\nonumber\\
e_1(\sqrt{g^{-1}f}) & = &\lambda_1 + \lambda_2 +\lambda_3 + \lambda_4 \\
e_2(\sqrt{g^{-1}f}) & = & \lambda_1\lambda_2 + \lambda_1\lambda_3 + \lambda_1\lambda_4 + \lambda_2\lambda_3
+ \lambda_2\lambda_4 + \lambda_3\lambda_4\nonumber\\
{\textbf{\ldots\ldots}} 
\end{eqnarray}

\section{Locality and nonlocality}

Once we have established the diffeomorphism symmetry of our model, the next step is to derive locality. 

According the Random Dynamics philosophy nature is inherently nonlocal, in field theory locality is however taken for granted, meaning that every degree of freedom is assigned a spatio-temporal site, i.e. that all interactions take place in one spacetime point. This implies that there is a system for assigning one site to each degree of freedom, and in a local theory the action can then be factorized. The partition function of the Universe then has the form
\begin{displaymath} Z=\int {\cal{D}}\psi e^{(iS+sources)} \end{displaymath} 	
where $S=S_1+S_2+...$, and each contribution only depends on the fields in limited regions of spacetime, corresponding to $S=\int {\cal{L}}(x) \mathrm{d}^4x$ in the continuum limit.

Nonlocality would then mean that a degree of freedom is a function of more than one spacetime point. An example of nonlocality is microcanonical ensemble, which in a formal sense is nonlocal - to approximate it to a canonical ensemble would from this perspective be analogous to approximating nonlocality with locality. In the microcanonical ensemble it is a constraint that gives rise to nonlocality, and this (omnipresent) nonlocality can be viewed as due to the presence of fixed extensive quantities, in a manner reminiscent of a microcanonical ensemble. This would then be a nonlocality inherent in nature, as opposed to one emerging from dynamical effects, i.e. not to the same as the ``nonlocality'' which refers to quantum nonlocality in the sense of non-separability, which occurs as nonlocal correlations which occur in settings such as the one discussed by Einstein, Podolsky and Rosen.  

\subsection{Fundamental nonlocality}
Since there are no instances in quantum mechanics of signals propagating faster than light, from the Random Dynamics point of view, quantum mechanics is not really nonlocal.
In the Random Dynamics scenario it is nonlocality that is taken for granted, locality appearing as a result of reparametrization invariance, i.e. as a result of diffeormorphism symmetry.
{\vspace{0.3cm}}

Our basic assumptions are as follows: 

\begin{itemize}
\item Locality only makes sense when you have a spacetime, or at least a manifold, so our starting point is a fundamental, differentiable manifold ${\cal{M}}$. To grant reparametrization invariance, we cannot do with simple Minkowski space, we
also need general relativity.
A reparametrization invariant formulation demands that also $g^{\mu\nu}$ gets transformed, since $g^{\mu\nu}$ = $\eta^{\mu\nu}$ would violate reparametrization invariance.
So if $g^{\mu\nu}$ is perceived as nothing but a field (i.e. in reality 10 fields), there is only a manifold.
Our manifold is moreover 4-dimensional, and it is only $g^{\mu\nu}$ that determines whether this means 4+0-dimensional, or $3+1$ - or $2+2$-dimensional.  
 
\item Some fundamental fields $\psi^k(x), A^{k\mu}(x),...,K^{k\mu\nu}(x)$,... defined on the manifold ${\cal{M}}$. 
We also want to have a $g^{\mu\nu}$ with contravariant, upper indices. Indices are important since upper and lower indices transform differently under reparametrization mappings, and if we were to include fermions, we should have vierbeins as well, presumeably with upper curved index. Assume that the chiral theory is formulated in terms of the Weyl equation, then we need vierbeins $e^\mu _a$ which transform as four-vectors with upper index, while $\psi$ transforms as a scalar under the curved index and thus reparametrization.
In addition there is flat index transformation under which $\psi$ transforms as a spinor, $e^\mu _a$ as a four-vector, and $g^{\mu\nu} $ as a scalar.

In higher dimensional theories you usually assume locality in the high dimensional space, for example in the case D=14,
$\int {\cal{L}}_4 \mathrm{d}^4x$
is local in higher dimensions. In an apriori arbitrary parametrization of the form 
${\cal{R}}^4 \mathrm {x} {\cal{R}}^{14-4}$, we get $\int {\cal{L}}_4 \mathrm{d}^4x$, where
\begin{equation} 
{\cal{L}}_4 \mathrm{d}^4x = \int {\cal{L}}(x,y) \mathrm{d}^{14-4}y
\end{equation} 
and ${\cal{L}}_4 \mathrm{d}^4x$ only depends on $\mathrm{x}$, while $\int {\cal{L}}(x,y) \mathrm{d}^{14-4}y$ only depends on ``infinitesimal'' neighbourhood in ($\mathrm{x},\mathrm{y}$); and in this sense the lower dimensions 'inherit' locality from the higher dimensions.  

Even if $\mathrm{y} \rightarrow \infty$ is non-compact far away, this argument is valid. That is, even in the case of non-compact extra dimensions, 4-locality is there.

\item Diffeomorphism symmetry, i.e. invariance under reparametrization mappings. 

Initially we however have a somewhat weaker assumption, demanding invariance only under 
$x \Rightarrow x'(x) = x + \epsilon(x)$, for det $(\partial x'^{\mu}/\partial x^{\nu}) = 1$.

\item We need some ``smoothness assumptions'', expecting Taylor expandability. 
When deriving locality we obviously don't start with a local action, so our starting function is just some generic action $S[g^{\mu \nu},\psi,\phi]$, where 
$\psi(x),\phi(x)$ are defined in four-dimensional spacetime represented by x, the reparametrization invariance implying that $S[\psi']= S[\psi]$.
\end{itemize}
For this action $S[g^{\mu \nu}, \psi, ...]$ we formulate some theorems:
\vspace{0.5cm}

$\underline{Theorem\hspace{2mm} I:}$

\noindent With our assumptions, the ``action'' $S[g^{\mu \nu}, \psi, ...]$ becomes a function of a basis for all the integrals you can form in a reparametrization invariant way from polynomials and mononomials in the fields and the derivatives at a single point x integrated over $\int ...\mathrm{d}^4x$ (i. e. the whole manifold).

{\vspace{0.2cm}}
\noindent We assume the manifold to be finite (compact), as a kind of infrared cutoff. 
\noindent Note that theorem I only implies a mild locality, i. e. an action of the form
\begin{equation}\label{local}
S = S(\int {\cal{L}}_1 \mathrm{d}^4x, \int {\cal{L}}_2 \mathrm{d}^4x,...).
\end{equation}
\noindent We derive something like a Lagrangian form, because we have many ${\cal{L}}_j$, and a complicated functional form.
\vspace{0.5cm}

$\underline{Theorem  \hspace{2mm}II:}$

When an action is of the form $S(\int {\cal{L}}_1 \mathrm{d}^4x, \int {\cal{L}}_2 \mathrm{d}^4x,...)$, called ``mild'' locality, then inside a small region of the manifold (a neighbourhood), and for a single field development, $g^{\mu\nu}_{actual},\psi_{actual} $, the ``Euler-Lagrange equations'' 

\begin{equation} 
\frac{\delta S}{\delta \psi(y)}|_{\psi=\psi_{actual},g^{\mu\nu}=g^{\mu\nu}_{actual}} = 0
\end{equation} 

\noindent are \underline{as if} the action were of the form $S = \int {\cal{L}}(x) \mathrm{d}^4x$ where ${\cal{L}}(x)$ is a linear combination of the ${\cal{L}}_j(x)'s$ with coefficients only depending on $g^{\mu\nu}_{actual}$ and  $\psi_{actual}$, but in such a way that these coefficients depend only very little on $g^{\mu\nu}_{actual},\psi_{actual}$ in the small local region considered.

According to Theorem II these ``coefficients'' do indeed exist, but it is apriori not certain that they are Taylor expandable. Actually there is a function-Taylor expansion for the function coming out of Theorem I.

\begin{equation} 
\psi^k(x)  \rightarrow  \psi^k(x)_{new}=\psi^k(x) \circ x'
\end{equation} 
for each fixed $k$, i.e. $\psi^k(x)_{new}=\psi^k(x'(x))=\psi^k(x)$, and
\begin{equation} 
A^{k\mu}_{new}(x'(x))=A^{k\nu}(x)\frac{\partial x'^{\mu}}{\partial x^{\nu}} {\hspace{4mm}}{\rm{and}}{\hspace{4mm}}
K^{k\rho \sigma}_{new}(x'(x))=K^{k\mu \nu}(x)\frac{\partial x'^{\rho}}{\partial x^{\mu}}\frac{\partial x'^{\sigma}}{\partial x^{\nu}}.
\end{equation} 
Proof of theorem I:
When we want to derive locality, we have to consider the ``locality postulates''. The first locality postulate is that
the Lagrangian ${\cal{L}}$ depends on an infinitesimal neighbourhood, i.e. $\int {\cal{L}} \mathrm{d}^4x$ is used for minimizing.
An evidently local action is then $S = \int {\cal{L}} \mathrm{d}^4x$, with ${\cal{L}} = {\cal{L}}(\psi, \partial\psi/\partial x,...)$; the goal being to formulate an action such that the
reparametrized action is a functional of the type 
\begin{equation} 
S(\psi') = {\cal{F}}(\int {\cal{L}}_1(x)\mathrm{d}^4x, \int {\cal{L}}_2(x)\mathrm{d}^4x,...,\int {\cal{L}}_n(x)\mathrm{d}^4x)
\end{equation} 

We also make the ``weak assumption'' that $S$ is functional expandable,
\begin{eqnarray} 
&&S[\psi] = \nonumber\\
&&\displaystyle\sum_{k=0}^{\infty} \int \int \int...\psi(x^{(1)})\psi(x^{(2)})...\psi(x^{(k)}) \frac{\delta S}{\delta \psi(x^{(1)}) \delta \psi(x^{(2)})...\delta \psi(x^{(k)})}\mathrm{d}^4x^{(1)}...\mathrm{d}^4x^{(k)}\nonumber\\
\end{eqnarray} 

The diffeomorphism symmetry implies that $S[\psi \circ x']=S[\psi]$, where $\psi' = \psi \circ x'$,
$\psi'(x) = \psi(x')=\psi(x'(x))$, the invariance meaning that $S[\psi']= S[\psi]$
In the Taylor expansion, one has to pay attention to that 
\begin{equation}
\frac{\delta \psi'(x)}{\delta \psi(y)} = \delta(x'(x)-y), 
\end{equation}
thus
\begin{equation}
\frac{\delta S[\psi']}{\delta \psi(y)}=\nonumber\\
\frac{\delta S[\psi(x'(x))]}{\delta \psi(y)}=\\
\int \frac{S[\psi]}{\delta \psi(y)}\delta(x' - y)\mathrm{d}^4x = det() \frac{\delta S[\psi]}{\delta \psi(x'^{-1}(y))}
\end{equation}

where we in the first round choose $det() = 1$.
Generalized:
\begin{equation}
\frac{\delta S[\psi']}{\delta \psi(y^{(1)})...\psi(y^{(k)})}=
det()\frac{\delta S[\psi]}{\delta \psi(x'^{-1}(y^{(1)}))...\delta \psi(x'^{-1}(y^{(k)}))}
\end{equation}

We want to choose $x'$ in such a way that $x'^{-1}(y^{(1)})=z^{(1)}, x'^{-1}(y^{(2)})=z^{(2)},..$, but with the demand that 
$z^{(j)} \neq y^{(k)}$ for all $j \neq k$.
If all $z^{(j)}$ are all different among themselves, and likewise the $y^{(j)}$ are all different among themselves, the functional derivative is a constant, but if we have a situation where some points are the same, e.g. $z_3=z_4=z_5$, 
the functional derivative will depend precisely on which points are not identical (under the reparametrizarion mapping that brings $z_3=z_4=z_5$ onto the points $y_3$, $y_4$, $y_5$, implying that $y_3=y_4=y_5$), i.e. 
$\delta S[\psi]/\delta \psi(y^{(1)})...\delta \psi(y^{(k)})$
only depends on how many in each group are identical.
All aberrances belong to a null set, and if we ignore this null set, we have
\begin{equation} 
\frac{\delta S[\psi]}{ \delta \psi(y^{(1)})...\delta \psi(y^{(k)})} = f_k
\end{equation} 
which is independent of the $y^{(j)}$'s. We then have 
\begin{equation} 
\begin{split} 
S[\psi] = \sum_{k=0}^{\infty} \frac{1}{k!} \int \cdots 
\int \frac{ \delta^k S}{\delta \psi(y^{(1)}
\cdots \delta \psi(y^{(k)}) }
 \psi(y^{(1)} \cdots \psi(y^{(k)}d^4y^{(1)} \cdots d^4 y^{(k)} = \\ 
= \sum \frac{f_k}{k!}  \int \cdots \int\psi(y^{(1)} \cdots \psi(y^{(k)} d^4y^{(1)} \cdots d^4 y^{(k)}
\end{split}
\end{equation} 
and 
\begin{equation} 
\sum_{k=0}^{\infty} \frac{f(k)}{k!}(\int\psi(y)\mathrm{d}^4y)^k = F(\int \psi(y)\mathrm{d}^4y),
\end{equation} 
so we got ``mild'' locality of the form (\ref{local}), i.e. some function of usual action-like terms
(in reality ``mild'' super local where super stands for no derivatives).

\noindent Now, if the null set argument is incorrect, consider that
\begin{equation} 
\frac{\delta S}{\delta\psi(y^{(1)})\delta\psi(y^{(2)})} = const. +\delta^4 (y^{(1)}-y^{(2)})
\end{equation} 
and  
\begin{eqnarray} 
&&\frac{\delta
  S}{\delta\psi(y^{(1)})\delta\psi(y^{(2)})...\delta\psi(y^{(k)})} =
\nonumber\\
&&C_1 + 
C_2 \displaystyle\sum_{j,l}^k \delta(y^{(j)} - y^{(k)}) + C_3 \sum \delta(y^{(j)} - y^{(k)}) \sum \delta(y^{(i)} - y^{(l)})
\end{eqnarray} 
where $C_j$ are constants.
Here we integrate over all points, whereby the same points might reappear several times. 
The resulting action is of the form
\begin{equation} 
S = F(\int \psi(x)\mathrm{d}^4x, \int \psi(x)^2 \mathrm{d}^4x,\int \psi(x)^3 \mathrm{d}^4x,...)
\end{equation} 
Now, what does such an action look like locally?

We can Taylor expand $S$:

\begin{equation}
\frac{\delta S[\psi]}{\delta \psi(y)}|_{\psi=\psi_a} =\nonumber\\
\displaystyle\sum_{\chi=1} \frac{\partial F}{\partial (\int\psi(x)^{\chi}\mathrm{d}^4x)} \chi \psi(x)^{\chi - 1}= f(\psi(x))
\end{equation}
where $\partial F/\partial (\int\psi(x)^{\chi}\mathrm{d}^4x) \chi\psi(x)^{\chi - 1}  $
can be locally approximated with a constant, and $f(\psi(x))$ depends on what happens in the entire universe.

We now have a situation where
$S \approx \int h(\psi(x)) \mathrm{d}^4x$ (where the function h is defined so that $h'(\psi) = f(\psi)$ i.e. it is the stem function of f), corresponding to a super local Lagrangian.

\subsection{An exercise}
As an exercise we will consider a theory with $\psi$ and $A^{\mu}$ (a contravariant vector field), keeping in mind that $A^{\mu}$ and $A_{\mu}$ transform differently under diffeomorphisms.

Taylor expanding the functional $S[\psi, A^{\mu}]$:
\begin{eqnarray}
&&S[\psi, A^{\mu}] = \nonumber\\
&&\displaystyle \sum_{k=0}^{\infty}\int\int\frac{\delta^k S}{\delta \psi(y^{(1)})} \delta \psi(y^{(2)})...\delta A^{\mu k}(y^{(k)})
\psi(y^{(1)})...A^{\mu k}(y^{(k)})\frac{1}{k!}\mathrm{d}^4y^{(1)}...\mathrm{d}^4y^{(k)}\nonumber\\
\end{eqnarray}

and consider
\begin{equation}
\frac{1}{1!}\int\frac{\delta S}{\delta A^{\mu}}(y^{(1)})\mathrm{d}^4y^{(1)}
\end{equation}
where $\delta S/\delta A^{\mu}(y^{(1)})$ is forced to be zero under reperametrization transformations.
But if we only include boundary terms,
  
\begin{equation}
\frac{\delta S}{\delta A^{\mu}(y^{(1)})} \sim \int \partial_{\mu}\delta(y^{(1)}-x)\mathrm{d}^4x \approx {\rm{only}}\hspace{2mm} {\rm{ boundary}}\hspace{2mm} {\rm{terms}}
\end{equation}
where the normal to the boundary $\eta_{\mu}$ $\sim$ $\partial f/\partial x^{\mu}$, and
\begin{equation}
\int_{\partial V_4}\eta_{\mu}A^{\mu}\mathrm{d}^3 x = \int A^{\mu}[dx]_{\mu},
\end{equation}

\begin{figure}[htb]
    \begin{center}
    \includegraphics[scale=0.7]{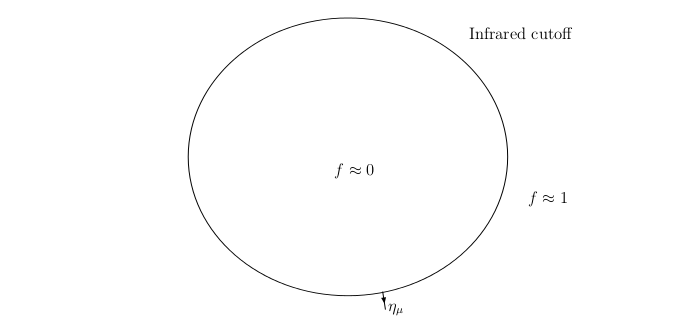}
\end{center}
    \end{figure}
the reparametrization invariance implies that
\begin{equation}
\frac{\delta S}{\delta A^{\mu}} = \eta_{\mu}\hspace{2mm}  {\rm{on}}\hspace{2mm}{\rm{the}}\hspace{2mm} {\rm{ boundary}},\hspace{2mm} {\rm{and}}\hspace{2mm}0\hspace{2mm}{\rm{on}}\hspace{2mm}{\rm{the}}\hspace{2mm}{\rm{inside}}\hspace{2mm}{\rm{of}}\hspace{2mm}V_4.
\end{equation}

\noindent
We want to have 
\begin{equation}
\int \frac{\delta S}{\delta A^{\mu}} A^{\mu}(y)\mathrm{d}^4y = \int_{\partial S}const. A^{\mu} \eta_{\mu}\mathrm{d}^3y|_{boundary}
\end{equation}

\noindent
This is integrated with $A^{\mu}$ as a variable, to
\begin{equation}
{\bf{C}}\delta_{\mu} A^{\mu} \mathrm{d}^4x = {\bf{C}} \int A_{\mu}\eta \mathrm{d}^3x 
\end{equation}
where ${\bf{C}}$ is a constant, and $\eta \mathrm{d}^3x$ represents the boundary.
\noindent Now the action is
\begin{equation}
 S = F(\int \psi(x)\mathrm{d}^4x, \int \psi(x)^2\mathrm{d}^4x,...,\int \partial_{\mu}A^{\mu} \mathrm{d}^4x, \int \psi(x)\partial_{\mu}A^{\mu} \mathrm{d}^4x,...)
\end{equation}
We now take all reparametrization invariant Lagrange density suggestions and let 
\begin{equation}
S = F(\int {\cal{L}}_1 \mathrm{d}^4x, \int {\cal{L}}_2 \mathrm{d}^4x,...)
\end{equation}
where we have remarked that the various integrands occuring (49), i.e. $\psi(x)$, $\psi(x)^2$ , ..., $\partial_{\mu}A^{\mu}(x) $, $\psi(x)\partial{\mu}A^{\mu}(x)$,...
are easily seen to be just those integrands
which ensures reparametrization invariance
(under our (simplifying) assumption of the determinant in the reparametrization $x'(x)$ being unity.).
We have therefore hereby finished the proof (or at least argument for) our above theorem I.

The theorem II is shown by arguing that, if we think of only investigating say the equations of motion in a small subregion of the whole spacetime region in which the universe have existed and will exist, then the integrals occurring in the function $F(\int {\cal L}_1(x) d^4x,
 (\int {\cal L}_2(x) d^4x, ...)$ will
only obtain a relatively very little part of their contribution for this very small local region. Thus these integrals as a whole will practically independent of the fields $\psi(x)$ etc. in the small region (where we live, and which is considered of interest). So indeed the statement of theorem II is true and we consider theorem II proven.

The final point is that we hereby have argue for that we {\em for practical purposes} got locality from assuming mainly diffeomorphism or reparametrization
invariance for practical purposes, in the sense that we only investigate it in an in space and time relative to the spacetime volume of the full existence of the universe small region. Further it were based on Taylor expandability of the very general a priori non-local action $S[\psi, A^{\mu}, ...]$.

This ``derivation'' of locality were initiated in collaboration with Don Bennett.

\section{Conclusion}

We have in this article sought to provide some - perhaps a bit speculative - ideas
for how to ``derive'' spacetime from very general starting conditions, namely a quantized analytical mechanical system. From a few and very reasonable assumptions, spacetime almost unavoidably appears, with the empirical properties of 3+1 dimensionality, reparametrization symmetry
- and thereby translational invariance, existence of fields, and practical locality (though not avoiding the nonlocalities due to quantum mechanics). Our initial assumption was that the states of the world were very close to a ground state, which in the phase space was argued to typically extend very far in $N$ dimensions, while only very shortly in the $N$ other dimensions. Here the number of degrees of freedom were called $N$ and thus the dimension of the phase $2N$. This picture of the ground state in the phase space we called the Snake,
because of its elongation in some, but not all directions. The long directions of the Snake becomes the protospace in
our picture. The translation and diffeomorphism symmetry are supposed to come about by first being formally introduced, but spontaneously broken by some ``Guendelmann fields $\xi$''. It is then argued that this spontaneous breaking is ``fluctuated away'' by quantum fluctuations, so that the symmetry truly appears, in the spirit of Lehto-Ninomiya-Nielsen. At the end we argued that once having gotten diffeomorphism symmetry, locality follows from simple Taylor expansion of the action and the diffeomorphism symmetry.

We consider this article as a very significant guide for how the project of Random Dynamics - of deriving all the known physical laws - could be performed in the range from having quantum mechanics and some smoothness assumptions to obtaining a useful spacetime manifold.

\section*{Acknowledgement}
First we would like to thank Don Bennett who initially was a coauthor of the last piece of this work: the derivation of locality; but then fell ill and could not finish and continue.

One of us (HBN) wants to thank the Niels Bohr Institute for support as an professor emeritus with office and support of the travels for the present work most importantly to Bled where this conference were held. But also Dr. Breskvar is thanked for economical support to this travel. 

We also thank for the discussions in Bled with our colleagues.

\end{document}